\documentclass[a4paper, 11pt]{article}
\usepackage[utf8]{inputenc}
\usepackage[english]{babel}
\usepackage{expl3}
\usepackage{array}
\usepackage{url}
\usepackage{amsmath,amsthm,amsfonts,amssymb,amscd}
\usepackage{fullpage}
\usepackage{algorithm}
\usepackage{algpseudocode}
\usepackage{tabularx}
\usepackage{booktabs}
\usepackage{lipsum}
\usepackage{bm}
\usepackage[colorlinks,urlcolor=cobalt,citecolor=cobalt,linkcolor=cobalt,pdftex, pdfauthor = Lukas Exl]{hyperref}
\usepackage{siunitx}
\usepackage{graphicx}
\usepackage{xcolor}
\usepackage{caption}
\usepackage{subcaption}
\usepackage[section]{placeins}
\usepackage{listings}
\usepackage{authblk}
\usepackage{stmaryrd} 

\usepackage{url}

\definecolor{cobalt}{rgb}{0.0, 0.28, 0.67}

\newtheorem{thm}{Theorem}
\newtheorem{lem}[thm]{Lemma}
\newtheorem{prop}[thm]{Proposition}

\theoremstyle{definition}

\DeclareUnicodeCharacter{2113}{\ensuremath{\ell}}

\renewcommand{\vec}[1]{\boldsymbol{#1}}

\newcommand*\change[1]{{\color{black}{#1}}}

\makeatletter

\newcounter{phase}[algorithm]
\newlength{\phaserulewidth}
\newcommand{\setphaserulewidth}{\setlength{\phaserulewidth}}

\makeatother
\setphaserulewidth{.7pt}

\DeclareSymbolFont{largesymbolsA}{U}{txexa}{m}{n}
\DeclareMathSymbol{\varprod}{\mathop}{largesymbolsA}{16}

\date{}

\title{Near-field-free super-potential FFT method for the three-dimensional free-space Poisson equation}
\author[a,b,c]{Lukas Exl \thanks{\texttt{lukas.exl@univie.ac.at}}}
\author[a,b]{Sebastian Schaffer}

\affil[a]{Math AI/ML, Wolfgang Pauli Institute, Vienna, Austria}
\affil[b]{Research Platform MMM Mathematics-Magnetism-Materials, University of Vienna, Vienna, Austria}
\affil[c]{Department of Mathematics, University of Vienna, Vienna, Austria}

\begin{document}
\maketitle
\noindent\textbf{Abstract.}
\noindent We present a spectrally accurate, efficient FFT-based method for the three-dimensional free-space Poisson equation with smooth, compactly supported sources. The method adopts a super-potential formulation: we first compute the convolution with the biharmonic Green’s function, then recover the potential by spectral differentiation, applying the Laplacian in Fourier space. A separable Gaussian-sum (GS) approximation enables efficient precomputation and quasi-linear, FFT-based convolution. Owing to the biharmonic kernel’s improved regularity, the GS cutoff error is fourth-order, uniform for all target points, eliminating the near-field corrections and Taylor expansions required in standard GS/Ewald-type methods. Benchmarks on Gaussian, oscillatory, and compactly supported densities reach the double-precision limit and, at matched accuracy on the same hardware, reduce both error and per-solve runtime relative to our original GS-based scheme. The resulting method is simple, reproducible, and efficient for three-dimensional free-space Poisson problems with smooth sources on uniform grids.



\section{Introduction}
This paper presents a method for solving the unbounded three-dimensional Poisson equation  
\begin{align}\label{eqn:problem1}
 - \Delta u(\boldsymbol{x}) = \rho(\boldsymbol{x}), \quad \boldsymbol{x} \in \mathbb{R}^3, \quad \lim_{|\boldsymbol{x}| \rightarrow \infty} |u(\boldsymbol{x})| = 0,
\end{align}
based on the classical representation of the solution as the convolution of the source density \(\rho\) with the free-space Green's function \(U(\boldsymbol{x}) = \tfrac{1}{4\pi} \tfrac{1}{|\boldsymbol{x}|}\),
\begin{align}\label{eqn:conv}
 u(\boldsymbol{x}) = (U \ast \rho)(\boldsymbol{x}) = \int_{\mathbb{R}^3} U(\boldsymbol{x} - \boldsymbol{y})\,\rho(\boldsymbol{y})\,d\boldsymbol{y}, \quad \boldsymbol{x} \in \mathbb{R}^3.
\end{align}

We assume the smooth and compactly supported density/source distribution \(\rho\) to be defined on a uniform grid, as it is common in many applications. This assumption enables the use of spectral methods based on trapezoidal quadrature, where the rapid decay of Fourier coefficients leads to high accuracy even on relatively coarse grids \cite{trefethen2000spectral,trefethen2014exponentially}.

The problem~\eqref{eqn:conv} is a fundamental model that appears in a wide range of physical applications \cite{campa2014physics}, including the simulation of Bose-Einstein condensates \cite{bao2015computing,jiang2014fast,bao2010efficient,mennemann2019optimal}, quantum chemistry~\cite{leach2001molecular,martyna1999rec,Wavelet06,Wavelet07,fusti2002accurate}, particle physics~\cite{arnold2013comparison,arnold2005efficient,hejlesen2016multiresolution}, electro- and magnetostatics \cite{guadagni2017fast,fisicaro2016generalized,exl_2014,exl_2014_nfft} and astrophysics~\cite{rampf2021cosmological,budiardja2011parallel}. 

Common approaches to compute the nonlocal convolution \eqref{eqn:conv} are based on Fourier transforms of the \textit{truncated Green's function} \cite{vico2016fast} or fall within the class of \emph{Ewald-type methods}~\cite{ewald1921berechnung,heyes1981electrostatic,martyna1999rec, exl2016accurate}, which decompose the singular Green's function \(U\) into a smooth long-range component \(U_s\) and a localized singular correction \(U_c\). This decomposition allows the smooth part of the convolution to be efficiently evaluated using the convolution theorem
\[
U \ast \rho \approx \mathcal{F}^{-1} \left( \mathcal{F}(U_s) \cdot \mathcal{F}(\rho) \right),
\]
typically on an equispaced Cartesian grid using the \textit{fast Fourier transform} (FFT), which offers quasi-linear scaling. The smoothness of \(U_s\) ensures rapid convergence of the Fourier series, making the spectral approximation both accurate and efficient. The remaining correction term involving \(U_c\) retains the singularity but is locally supported, and thus can be evaluated using direct summation or FFT-based correction via a Taylor expansion of the density \cite{exl2016accurate,exl2017gpu}. A naive direct evaluation of the full convolution~\eqref{eqn:conv} scales as \(\mathcal{O}(N^2)\) on a grid with \(N\) points. In contrast, the original Ewald method~\cite{ewald1921berechnung} and optimized variations scale as \(\mathcal{O}(N^{3/2})\). A fully spectral method as introduced in~\cite{exl2016accurate} is based on a smooth and separable \textit{Gaussian-sum} (GS) approximation of the Green’s function \cite{hackbusch2006low,beylkin2010approximation,exl_2014,greengard2018anisotropic} to regularize the singularity, that is,
\begin{align}
    U(\vec x) \approx U_{GS}(\vec x) = \sum_{s=0}^S \omega_s \prod_{p=1}^3 e^{-\alpha_s x_p^2}, \quad |\vec x| \in (\varepsilon, \delta],\, 0 < \varepsilon < \delta. 
\end{align}
Owing to the tensor-product structure of the approximation, a significant portion of the computational effort can be accurately precomputed in the form of one-dimensional integrals during the setup phase, where the effort is affordable since problem \eqref{eqn:conv} is typically solved many times in applications. However, the cutoff error introduced by the Gaussian-sum approximation is of the order \(\mathcal{O}(\varepsilon^2)\), which requires the evaluation of a correction integral to maintain high overall accuracy. In the original GS-based method~\cite{exl2016accurate}, this correction was performed using a Taylor expansion of the density function followed by FFT evaluation, introducing additional computational effort.

In contrast, the method proposed in this work achieves \(\mathcal{O}(N \log N)\) complexity by applying FFT-based convolution only to a smooth component of the biharmonic Green's function, eliminating the need for a correction term while maintaining accuracy up to machine precision.

The key component of the method is the introduction of a \textit{super-potential} \(v\) \cite{chandrasekhar1989super}, which satisfies the simple relation  
\begin{align}\label{eqn:superpot}
    u(\boldsymbol{x}) = \Delta v(\boldsymbol{x}),
\end{align}
where \(v\) solves the biharmonic equation $-\Delta^2 v = \rho$ and is given as the convolution with the biharmonic Green’s function. After computing \(v\) via Fourier-based convolution, the solution \(u\) to~\eqref{eqn:conv} can then be efficiently obtained through spectral differentiation, maintaining high precision. We approximate the convolution with the biharmonic kernel using a rank-3 separable Gaussian-sum resulting in a truncation error of order \(\mathcal{O}(\varepsilon^4)\) \cite{exl2025higher}. \change{The original GS-based method \cite{exl2016accurate} yields an error for the smooth convolution -- that is without near-field correction -- of $\mathcal{O}(\varepsilon^2)$}. \change{Relative to the original GS scheme -- which requires a localized Taylor correction of the source -- our super-potential formulation removes the near-field stage entirely while delivering a uniform fourth-order cutoff error for all targets, simplifying parameters and improving per-solve efficiency at high accuracy.}

\section{Method}
For the sake of simplicity, we assume a (rescaled) density compactly supported in the unit cube $\boldsymbol{B}_1 = [-1/2,1/2]^3$ (a general rectangular domain could also be considered \cite{exl2017gpu}). The problem \eqref{eqn:conv} gets for $\vec x \in \boldsymbol{B}_1$
\begin{align}\label{eqn:conv2}
 u(\boldsymbol{x}) = (U \ast \rho)(\boldsymbol{x}) = \int_{\mathbb{R}^3} U(\boldsymbol{y})\,\rho(\boldsymbol{x}-\boldsymbol{y})\,d\boldsymbol{y} = \int_{\boldsymbol{B}_2} U(\boldsymbol{y})\,\rho(\boldsymbol{x}-\boldsymbol{y})\,d\boldsymbol{y}
\end{align}
with $\boldsymbol{B}_2 := 2 \cdot \boldsymbol{B}_1 = [-1,1]^3$.
According to \eqref{eqn:superpot} we have \change{for the biharmonic kernel $V(\vec y) = \tfrac{1}{8\pi}\,|\vec y|$}
\begin{align}\label{eqn:superpot1}
    u(\vec x) = \change{\Delta v(\vec x) = \, } \Delta \big(V \ast \rho\big)(\vec x) = \Delta \int_{\boldsymbol{B}_2} \Big(\frac{1}{8\pi}\Big) |\boldsymbol{y}| \,\rho(\boldsymbol{x} - \boldsymbol{y})\, d\boldsymbol{y}.
\end{align}
After a Fourier series approximation of the density, we can apply the convolution theorem utilizing FFT. Note that $\vec x - \vec y \in 3\cdot \boldsymbol{B}_1$. However, it is enough to consider a (truncated) Fourier series approximation of $\rho$ on $\boldsymbol{B}_2$ (with simple zero padding factor 2) \cite{exl2016accurate}.  More precisely, let the $2$-periodic ($L=2$) smooth density $\rho$ be approximated as 
\begin{align}
\rho(\boldsymbol{x}) \approx \sum_{\boldsymbol{k}} \hat{\rho}_{\boldsymbol{k}} \, e^{2\pi i\, \frac{\boldsymbol{k} \cdot \boldsymbol{x}}{L}},
\end{align}
$\boldsymbol{k} = (k_1,k_2,k_3)$, $k_j = -\tilde{n}_j/2,\hdots,\tilde{n}_j/2-1,\, \tilde{n}_j = 2 n_j$  with the Fourier coefficients  
\begin{align}
\hat{\rho}_{\boldsymbol{k}} = \frac{1}{8} \int_{\boldsymbol{B}_2} \rho(\boldsymbol{x}) \, e^{-2\pi i\, \frac{\boldsymbol{k} \cdot \boldsymbol{x}}{L}} \, d\boldsymbol{x},
\end{align}
which are approximated by trapezoidal quadrature utilizing FFT.
After substitution into \eqref{eqn:superpot1} \change{and spectral differentiation} we get
\begin{align}\label{eqn:convtheorem}
u(\boldsymbol{x}) = -\pi^2 \sum_{\boldsymbol{k}} ( |\boldsymbol{k}|^2\, \hat{V}_{\boldsymbol{k}} \, \hat{\rho}_{\boldsymbol{k}}) \, e^{\pi i\, \boldsymbol{k} \cdot \boldsymbol{x}}
\end{align}
with \change{the smooth integrals}
\begin{align}\label{eqn:Gten0}
\hat{V}_{\boldsymbol{k}} := \int_{\boldsymbol{B}_2} \Big(\frac{1}{8\pi}\Big) |\boldsymbol{y}| \, e^{-\pi i\, \boldsymbol{k} \cdot \boldsymbol{y}} \, d\boldsymbol{y},
\end{align}
where \eqref{eqn:convtheorem} is computed by (inverse) FFT.

In principle, the integrals in~\eqref{eqn:Gten0} could be precomputed. However, accurate evaluation of the three-dimensional integrals is both computationally expensive and numerically challenging. To overcome this and efficiently shift a significant portion of the computational cost to the setup phase, we employ a separable Gaussian-sum approximation of the kernel \( |\boldsymbol{y}| \). This enables effective and accurate precomputation by reducing the problem to sums of one-dimensional tensor-product integrals.

\begin{figure}
\center
\includegraphics[width=0.75\textwidth]{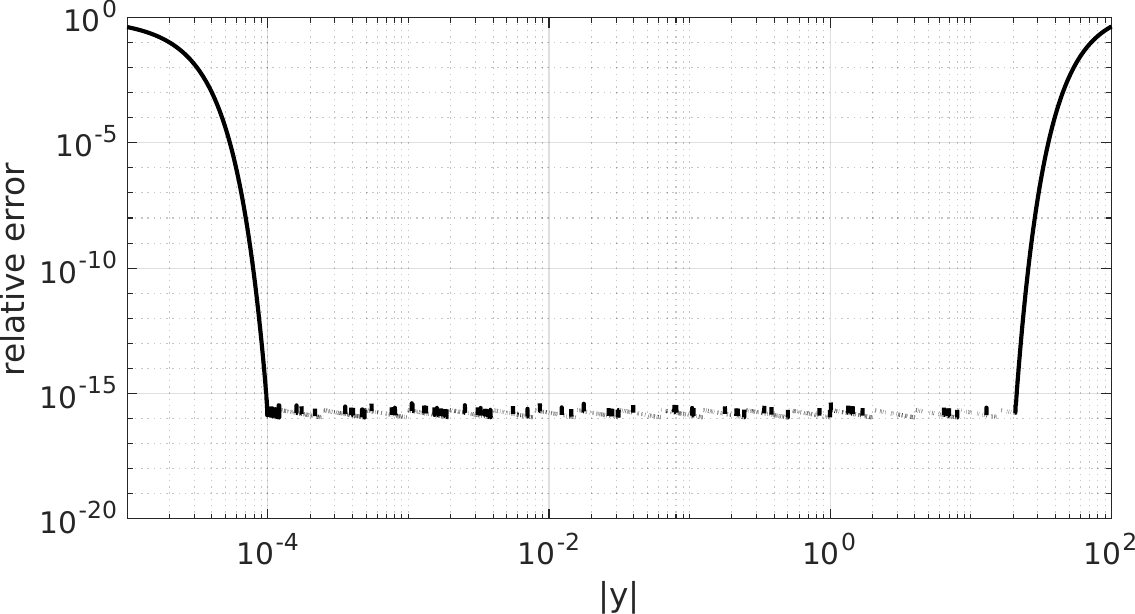}
\caption{\label{fig:GSapprox}Relative errors of the GS approximation for $|y|$ with $S=460$ terms.}
\end{figure}

We begin with the integral representation
\begin{align}
 \int_0^\infty t^n e^{-r\, t^2}\,\text{d}t = \frac{\Gamma \left(\tfrac{n+1}{2}\right)}{2 r^{\tfrac{n+1}{2}}}, \quad r > 0,\quad n > -1,
\end{align}
and consider the case \(n = 0\). Following the \textit{sinc quadrature} approach \cite{hackbusch2006low}, this yields an exponentially convergent, separable approximation of $\tfrac{1}{8\pi}\tfrac{1}{|\boldsymbol{y}|}$ for \(r = |\boldsymbol{y}|^2\). Specifically, for \(\boldsymbol{y} \in \mathbb{R}^3\) with \(0 < \varepsilon < 1 \leq \delta\), we obtain
\begin{align}\label{eq:sin_1_x}
    \Big(\frac{1}{8\pi}\Big)\frac{1}{|\boldsymbol{y}|} \approx \sum_{s=0}^S \omega_s \, e^{-\alpha_s |\boldsymbol{y}|^2} 
    = \sum_{s=0}^S \omega_s \prod_{p=1}^3 e^{-\alpha_s y_p^2}, \quad |\boldsymbol{y}| \in (\varepsilon, \delta],
\end{align}
where the approximation is given as a sum of separable Gaussians.

Multiplying both sides of~\eqref{eq:sin_1_x} by \(|\boldsymbol{y}|^2\) yields an approximation of the biharmonic kernel
\begin{align}\label{eq:gs_x}
    V(\boldsymbol{y}) \approx V_{\mathrm{GS}}(\boldsymbol{y}) 
    = \sum_{s=0}^S \omega_s |\boldsymbol{y}|^2 e^{-\alpha_s |\boldsymbol{y}|^2} 
    = \sum_{s=0}^S \sum_{q=1}^3 \omega_s 
    \prod_{p=1}^3 
    \begin{cases}
        y_p^2\, e^{-\alpha_s y_p^2}, & \text{if } p = q, \\[0.5ex]
        e^{-\alpha_s y_p^2}, & \text{if } p \neq q,
    \end{cases}
\end{align}
which is a sum of rank-3 separable functions. The sinc quadrature parameters and the approximation interval \((\varepsilon, \delta]\) can be adapted based on the desired accuracy and domain size~\cite{exl2016accurate,exl_2014}. Figure~\ref{fig:GSapprox} illustrates the relative error of the GS approximation \change{obtained from sinc quadrature with \(S=460\)} terms and step parameter $c_0=1.9$, showing double-precision accuracy in $(10^{-4},\delta]$ with $\delta > 1$.

In the super-potential convolution, \change{$v(\vec x) = (V \ast \rho)(\vec x)$}, in \eqref{eqn:superpot1}, replacing the kernel \(|\boldsymbol{y}|\) with \(V_{\mathrm{GS}}(\boldsymbol{y})\) results in an overall approximation error of order \(\mathcal{O}(\varepsilon^4)\), provided that the pointwise difference \(|\,\frac{1}{8\pi}|\boldsymbol{y}| - V_{\mathrm{GS}}(\boldsymbol{y})\,|\) is negligible in the interval \((\varepsilon, \delta]\) \change{away from the origin}. \change{Hence, the remainder is localized  as follows:}
\begin{align}
    \Big(\int_{\boldsymbol{B}_2\setminus \mathcal{B}_\varepsilon(\vec 0)} +  \int_{\mathcal{B}_\varepsilon(\vec{0})} \Big)\Big(\Big(\frac{1}{8\pi}\Big)|\vec{y}| - V_{\mathrm{GS}}(|\vec{y}|)\Big)\, \rho(\vec{x}-\vec{y})\, d\boldsymbol{y}
    \approx 
    \int_{\mathcal{B}_\varepsilon(\vec{0})} \Big(\Big(\frac{1}{8\pi}\Big)|\vec{y}| - V_{\mathrm{GS}}(|\vec{y}|)\Big)\, \rho(\vec x - \vec{y})\, d\boldsymbol{y},
\end{align}
where \(\mathcal{B}_\varepsilon(\vec{0})\) denotes the ball of radius \(\varepsilon\) centered at the origin. \change{The following Lemma shows that replacing $V$ by $V_{GS}$ yields an absolute error of $\mathcal{O}(\varepsilon^4)$ uniformly in the target $\vec x$ -- that is, for all target points in $\boldsymbol{B}_1$, including the near-field.}

\begin{lem}\label{lem:correctionint}
Let $\varepsilon > 0,\, \rho$ be bounded, and $|\,\frac{1}{8\pi}|\boldsymbol{y}| - V_{\mathrm{GS}}(\boldsymbol{y})\,| = \mathcal{O}(\varepsilon^4)$ for $|\vec y| \in (\varepsilon,1]$. Then there holds \change{for $\vec x \in \boldsymbol{B}_1$}
\begin{align}
    \change{v(\vec x) =} \int_{\boldsymbol{B}_2} \frac{1}{8 \pi} |\boldsymbol{y}| \,\rho(\boldsymbol{x} - \boldsymbol{y})\, d\boldsymbol{y} = \int_{\boldsymbol{B}_2} V_{GS}(\boldsymbol{y}) \,\rho(\boldsymbol{x} - \boldsymbol{y})\, d\boldsymbol{y} + \mathcal{O}(\varepsilon^4).
\end{align}
\end{lem}
\begin{proof} We have 
\begin{align}
    \int_{\boldsymbol{B}_2} \Big(\Big(\frac{1}{8\pi}\Big)|\vec{y}| - V_{\mathrm{GS}}(|\vec{y}|)\Big)\, \rho(\vec{x}-\vec{y})\, d\boldsymbol{y}
    = 
    \Big(\int_{\boldsymbol{B}_2\setminus \mathcal{B}_\varepsilon(\vec{0})} + \int_{\mathcal{B}_\varepsilon(\vec{0})}\Big) \Big(\Big(\frac{1}{8\pi}\Big)|\vec{y}| - V_{\mathrm{GS}}(|\vec{y}|)\Big)\, \rho(\vec x - \vec{y})\, d\boldsymbol{y},
\end{align}
where due to the accuracy of the GS approximation 
\begin{align}
    \Big| \int_{\boldsymbol{B}_2\setminus \mathcal{B}_\varepsilon(\vec{0})} \Big(\Big(\frac{1}{8\pi}\Big)|\vec{y}| - V_{\mathrm{GS}}(|\vec{y}|)\Big)\, \rho(\vec x - \vec{y})\, d\boldsymbol{y} \Big| \leq |\boldsymbol{B}_2| \,\|\rho\|_\infty \, \max_{|\vec y| \in (\varepsilon,1]} |\,\Big(\frac{1}{8\pi}\Big)|\boldsymbol{y}| - V_{\mathrm{GS}}(\boldsymbol{y})\,| = \mathcal{O}(\varepsilon^4).
\end{align}

For $\boldsymbol{y} \in \mathcal{B}_{\varepsilon}(\vec 0)$ we have $|\vec y| \leq \varepsilon$ and $V_{GS}(|\vec y|) \leq C \varepsilon^2 + \mathcal{O}(\varepsilon^4)$ and hence,
\begin{align}
    \Big|\int_{\mathcal{B}_{\varepsilon}(\vec 0)} \Big(\Big(\frac{1}{8\pi}\Big)|\vec y|- V_{GS}(|\vec y|)\Big) \,\rho(\vec x - \vec y)\, d\boldsymbol{y} \Big| \leq \varepsilon \,\Big(\frac{1}{8\pi}+C\, \varepsilon\, + \mathcal{O}(\varepsilon^3)\Big) \, |\mathcal{B}_{\varepsilon}(\vec 0)| \, \|\rho\|_\infty= \mathcal{O}(\varepsilon^4), 
\end{align}
with $|\mathcal{B}_{\varepsilon}(\vec 0)| = \mathcal{O}(\varepsilon^3)$ denoting the volume of the $\varepsilon$-ball.
\end{proof}

The approximations of the integrals \eqref{eqn:Gten0} can now be precomputed as a sum of tensor products of one-dimensional  quadratures (e.g. Gauss-Kronrod as in the original GS-based method \cite{exl2016accurate}) up to machine precision. In fact, we have
\begin{align}\label{eqn:Gten_sep}
\hat{V}_{\boldsymbol{k}} \approx \int_{\boldsymbol{B}_2} V_{GS}(\boldsymbol{y}) \, e^{-\pi i\, \boldsymbol{k} \cdot \boldsymbol{y}} \, d\boldsymbol{y} = 
\sum_{s=0}^S \sum_{q=1}^3 \omega_s 
    \prod_{p=1}^3 
    \begin{cases}
        \int_0^1 2\, y_p^2\, e^{-\alpha_s y_p^2} \, \cos(\pi k_p y_p)\, dy_p, & \text{if } p = q, \\[0.15cm]
        \int_0^1 2\, e^{-\alpha_s y_p^2}\,\cos(\pi k_p y_p)\, dy_p , & \text{if } p \neq q.
    \end{cases}
\end{align}
In general \eqref{eqn:Gten_sep} \change{only} needs the computation \change{and storage} of $2 \times (n_1 + n_2 + n_3) \times (S+1)$ one-dimensional integrals with $n_j = \tilde{n}_j/2$.
In contrast, the tensor \eqref{eqn:Gten0} would need the accurate computation of $\mathcal{O}(n_1 n_2 n_3)$ three-dimensional integrals.


Our method consists of a simple FFT-based convolution of the density $\rho$ with the precomputed tensor components $\hat{V}_{\boldsymbol{k}}$ for the super-potential $v$ with cutoff rate $\mathcal{O}(\varepsilon^4)$ and application of the spectral Laplacian to obtain the potential $u$ according to \eqref{eqn:superpot}., which preserves the cutoff rate.

\begin{prop}[Spectral Laplacian preserves the cutoff rate, \cite{trefethen2000spectral}]
Let $v_h$ be the grid function obtained by the GS/FFT convolution and $u_h:=\Delta_h v_h$ the result of applying the Laplacian spectrally (i.e., multiplying resolved Fourier modes by $-|\vec k|^2$ on the padded grid). Then, for fixed grid size $N$ and padding, there exists a constant $C(N)$ depending only on the maximal resolved wavenumber such that
\[
\|u-u_h\|_{\ell^\infty} \le C(N)\,\|v-v_h\|_{\ell^\infty}.
\]
In particular, since $\|v-v_h\|_{\ell^\infty}=O(\varepsilon^4)+(\text{spectral in }N)$ for smooth data, we have
\[
\|u-u_h\|_{\ell^\infty} = O(\varepsilon^4) + (\text{spectral in }N).
\]
\end{prop}

\begin{proof}
On the padded grid, the spectral Laplacian is the diagonal Fourier multiplier $-|\vec k|^2$; in physical space this defines a bounded linear operator on the resolved trigonometric polynomial space. Hence
$\|u-u_h\|_{\ell^\infty}=\|\Delta(v-v_h)\|_{\ell^\infty}\le C(N)\|v-v_h\|_{\ell^\infty}$,
where $C(N)$ can be taken as the maximal multiplier magnitude mapped back to $\ell^\infty$ via the inverse FFT norm. The stated rate follows immediately.
\end{proof}



Combining the GS approximation on $(\varepsilon,\delta]$, the near-ball estimate, and the spectral application of the Laplacian yields the following a priori bound for the potential $u$.

\begin{thm}[A priori error bound for the potential $u$]\label{thm:u-bound}
Let $U(r)=\tfrac{1}{4\pi r}$ and $V(r)=\tfrac{1}{8\pi}r$ denote the Laplace and biharmonic Green’s functions in three dimensions.
Assume a GS approximation $U_{\mathrm{GS}}$ satisfies
\[
\max_{r\in(\varepsilon,\delta]} |U(r)-U_{\mathrm{GS}}(r)| \;\le\; \eta_S .
\]
Define $V_{\mathrm{GS}}(r):=r^2\,U_{\mathrm{GS}}(r)$ and let
\[
v := V * \rho, \qquad v_h := V_{\mathrm{GS}} * \rho ,
\]
with $\rho\in C_0^\infty(\boldsymbol{B}_1)$ supported in the unit box $\boldsymbol{B}_1$. Let $u:=\Delta v$ and let
$u_h$ be obtained by applying the spectral Laplacian on the padded grid (FFT multiplier by $-|\vec k|^2$) to $v_h$ and cropping back to $B_1$.
Then, for fixed grid size and padding, there exists a constant $C_{\Delta}(N)$ depending only on the maximal resolved wavenumber such that
\[
\|u-u_h\|_{\ell^\infty(\boldsymbol{B}_1)}
\;\le\;
C_{\Delta}(N)\,\Big(
\underbrace{\delta^{2}\,\eta_S\,\|\rho\|_{L^1(\boldsymbol{B}_1)}}_{\text{outer region }(\varepsilon,\delta]}
\;+\;
\underbrace{C_0\,\varepsilon^{4}\,\|\rho\|_{\infty}}_{\text{inner ball }[0,\varepsilon]}
\;+\;
\underbrace{R_{\mathrm{tail}}(\delta)}_{\text{truncation beyond }\delta}
\Big).
\]
In particular, choosing $\eta_S=O(\varepsilon^{4})$ yields
\[
\|u-u_h\|_{\ell^\infty(\boldsymbol{B}_1)} \;=\; O(\varepsilon^{4}) \;+\; \text{(spectral in $N$ for smooth data),}
\]
uniformly for all target points $x\in \boldsymbol{B}_1$.
\end{thm}

\begin{proof}
First, as in the convolution bound, split the error for $v$ into contributions from $(\varepsilon,\delta]$, from $[0,\varepsilon]$, and from the truncated tail $(\delta,\infty)$:
on $(\varepsilon,\delta]$ we have $|V-V_{\mathrm{GS}}|=r^2|U-U_{\mathrm{GS}}|\le \delta^2\eta_S$,
the near-ball contribution scales like $\int_{\mathcal{B}_\varepsilon(\vec 0)} |y|\,dy = O(\varepsilon^4)$,
and the tail is $R_{\mathrm{tail}}(\delta)$ by construction of the window.
Hence
$\|v-v_h\|_{\ell^\infty(\boldsymbol{B}_1)}\le \delta^2\eta_S\|\rho\|_{L^1}+C_0\varepsilon^4\|\rho\|_{C^0}+R_{\mathrm{tail}}(\delta)$.
Applying the spectral Laplacian multiplies each resolved Fourier mode by $-|\vec k|^2$; as a bounded linear operator on the resolved trigonometric space, it satisfies
$\|u-u_h\|_{\ell^\infty}\le C_{\Delta}(N)\,\|v-v_h\|_{\ell^\infty}$,
which gives the stated bound. For smooth (analytic) $\rho$, the grid-dependent part decays spectrally with $N$.
\end{proof}

\section{Numerical results}
In the following the method is validated for the benchmarks from \cite{exl2017gpu} using several choices of the density with known exact solutions, where \change{we start with a Gaussian distribution as source on different domains and also show for varying near-field cutoff $\varepsilon$ the uniform error in the target locations in the order of $\mathcal{O}(\varepsilon^4)$ according to Lemma~\ref{lem:correctionint}. In all the other numerical tests} we use the \change{(standard)} GS approximation illustrated in Fig.~\ref{fig:GSapprox} with $\varepsilon = 10^{-4}$. We calculate the maximum relative errors $E$ according to 
\begin{align}\label{error_infty}
E:=\frac{\|u-u_{\vec h}\|_{\ell^\infty}}{\|u\|_{\ell^\infty}}=
\frac{\max_{\vec x\in \mathcal{T}_h} |u(\vec x)-u_{\vec h}(\vec x)|}{\max_{\vec x\in \mathcal{T}_h} |u(\vec x)|},
\end{align}
where $\mathcal{T}_h$ is the rectangular computational domain discretized uniformly in each direction with mesh sizes $\vec h = (h_x,h_y,h_z)$. \change{The number of total grid points is denoted with $N = n_xn_yn_z$, with $(n_x,n_y,n_z)$ being the number of grid points in each direction.}
For the numerical experiments we use MATLAB on a notebook with Intel Core i7-8565U.
We perform all computations with double precision. 
The \change{CPU} timings \change{$t$ for the convolution} are measured as the average times of $10$ computations. \change{Note that the precomputation of the integrals \eqref{eqn:Gten_sep} takes place in a once-for-all setup phase for a given Cartesian grid, lasting about 100 seconds in the double-precision limit case of a near-field cutoff of $\varepsilon = 10^{-4}$ and a grid size of $N=128^3$. 
Timings report only the solve phase; the offline precomputation -- building the Gaussian-sum tables, generating FFT plans, and caching the 1D integrals -- is performed once per ($\varepsilon$, domain, grid) configuration (typically fixed within a simulation) and is reused for all right-hand sides.}
We first give an algorithm summary in the next subsection.

\subsection{Algorithm summary}
Our method is summarized in the pseudo-codes Alg.~\ref{alg:precompute} and Alg.~\ref{alg:solve} as a one-time precomputation and the actual solve routine, respectively. 

\begin{algorithm}[H]
\caption{One-time precomputation (per $(\varepsilon,\text{domain},\text{grid})$ configuration)}
\label{alg:precompute}
\begin{algorithmic}[1]
\Require cutoff $\varepsilon$ (default $1$E-$04$), GS rank $S$, sinc step $c_0$ (default $1.9$), Gauss--Kronrod tolerance $\mathtt{tol}_{\mathrm{GK}}$ (default machine precision); grid sizes $(n_x,n_y,n_z)$ and spacings $(h_x,h_y,h_z)$; padding factor $p \equiv 2$
\Ensure
\State Compute weights/nodes $(\omega_s,\alpha_s),\, s=0,\hdots,S$ for GS approximation $V_{GS}$ of $V$ via \eqref{eq:gs_x} 
\State Compute tensor components $\hat{V}_{\boldsymbol{k}}$ via \eqref{eqn:Gten_sep}
\State Establish FFT plans on the padded grid
\end{algorithmic}
\end{algorithm}

\begin{algorithm}[H]
\caption{Solve for $u$ given a source $\rho$ (super-potential GS/FFT pipeline)}
\label{alg:solve}
\begin{algorithmic}[1]
\Require Samples $\rho$ on the $(n_x,n_y,n_z)$ grid; precomputation of tensor components $\hat{V}_{\boldsymbol{k}}$ and FFT plans via Alg.~\ref{alg:precompute} 
\Ensure Approximate solution $u$ on the original grid
\State Compute Fourier coefficients $\hat{\rho}_{\boldsymbol{k}}$ of density $\rho$ via FFT on padded grid
\State Multiply $( |\boldsymbol{k}|^2\, \hat{V}_{\boldsymbol{k}} \, \hat{\rho}_{\boldsymbol{k}})$ (see \eqref{eqn:convtheorem})
\State Compute $u$ in \eqref{eqn:convtheorem} via inverse FFT and crop back to original $(N_x,N_y,N_z)$ grid
\State \Return $u$
\end{algorithmic}
\end{algorithm}

\subsection{Benchmarks on Gaussian sources}
First we test with the Gaussian density 
\begin{align}\label{eqn:gaussian}
 \rho(\vec x) = \frac{1}{(2\pi)^{3/2}\sigma^3 } e^{-|\vec x-\vec c|^2/(2\sigma^2)}, 
\end{align}
where $\vec c \in \mathbb{R}^3$ is the center of the computational box and the exact solution is 
\begin{align}
 u^\ast(\vec x) = \frac{1}{4 \pi |\vec x-\vec c|} \textrm{Erf}\Big(\frac{|\vec x-\vec c|}{\sqrt{2}\sigma}\Big), 
\end{align}
\change{where Erf denotes the error function.} We compute the potential in the computational domain $\boldsymbol{B}=[-2,2]^3$ with a varying shape parameter $\sigma$ and compare errors and computation times; see Tab.~\ref{tab:gaussian}. For our proposed method we observe up to exponentially fast convergence \change{to the double-precision limit} and $\mathcal{O}(N\log N)$ scaling. Compared to the original GS-based method, we need no near-field correction via Taylor expansion of the density and one further FFT calculation.

\change{In addition, in Tab.~\ref{tab:GSorig} we show the respective results for the original GS-based method of \cite{exl2016accurate,exl2017gpu} \textit{with and without} near-field correction. Omitting the near-field correction in the original GS-based method leads to a theoretical error in the order of $\mathcal{O}(\varepsilon^2)$. For fairness, we use identical parameters for the sinc quadrature and the Gauss–Kronrod rule as for the computations in Tab.~\ref{tab:gaussian}, and the same near-field cutoff $\varepsilon=10^{-4}$. The baseline code of \cite{exl2017gpu} (MATLAB) is modified solely to match these settings and executed on the same CPU. The solve times in the case of no near-field correction are comparable to our method, since in both cases the same amount of FFTs of the same size are used, whereas speedups \change{of about 20-25\%} compared to the original method presented in \cite{exl2017gpu} \change{including the near-field correction can be observed. The precomputation times are less compared to our proposed method, that is, about 25 sec for a grid size $N=128^3$ vs. 100 sec for our method. The presented errors in Tab.~\ref{tab:GSorig} confirm the need of a near zone correction in the original GS method if very high accuracy is needed.}}

\begin{table}[h!]
\caption{Errors $E$ and timings $t$ (s) for the Gaussian density \eqref{eqn:gaussian} in $[-2,2]^3$ \change{with $\varepsilon=10^{-4}$ calculated with our proposed method.}} 
\label{tab:gaussian}
\centering
\setlength{\tabcolsep}{16pt}
\begin{tabular}{c c c c}
\toprule
$\sigma$ & $N$     & $E$           & $t$ \\ \hline
$0.20$   & $16^3$  & 1.659E-03  & 1.37E-03 \\      
$0.20$   & $32^3$  & 4.154E-09  & 5.13E-03 \\
$0.20$   & $64^3$  & 5.555E-15  & 4.81E-02 \\
$0.20$   & $128^3$ & 3.859E-15  & 5.51E-01 \\\hline

$0.15$   & $16^3$  & 2.986E-02   & 7.26E-04 \\
$0.15$   & $32^3$  & 2.937E-06   & 4.85E-03 \\
$0.15$   & $64^3$  & 8.917E-15   & 6.05E-02 \\
$0.15$   & $128^3$ & 5.077E-15   & 5.36E-01 \\\hline

$0.10$   & $16^3$  & 3.802E-01   & 6.96E-04 \\
$0.10$   & $32^3$  & 1.129E-03   & 4.91E-03 \\
$0.10$   & $64^3$  & 2.624E-09   & 4.90E-02 \\
$0.10$   & $128^3$ & 8.232E-15   & 5.35E-01 \\
\bottomrule
\end{tabular}
\end{table}

\begin{table}[h!]
\caption{\change{Errors $E$ and timings $t$ (s) for the Gaussian density \eqref{eqn:gaussian} in $[-2,2]^3$ \change{with $\varepsilon=10^{-4}$} calculated with the original GS-based method of \cite{exl2016accurate,exl2017gpu}. For comparison, the values in brackets in the third and fourth column represent the errors and timings of the original GS-based method \textit{without} near-field correction.}} 
\label{tab:GSorig}
\centering
\setlength{\tabcolsep}{16pt}
\begin{tabular}{c c c c}
\toprule
$\sigma$ & $N$     & $E$           & $t$ \\ \hline
$0.20$   & $16^3$  & 1.659E-03 \, (1.659E-03)  & 2.82E-03 \, (1.48E-03) \\     
$0.20$   & $32^3$  & 4.154E-09 \, (9.298E-09)  & 6.16E-03 \, (4.99E-03) \\
$0.20$   & $64^3$  & 7.082E-16 \, (7.405E-09)  & 7.12E-02 \, (5.85E-02)\\
$0.20$   & $128^3$ & 1.403E-15 \, (7.429E-09)  & 6.90E-01 \, (5.78E-01)\\\hline

$0.15$   & $16^3$  & 2.986E-02 \, (2.986E-02)   & 8.64E-04 \, (7.22E-04) \\
$0.15$   & $32^3$  & 2.937E-06 \, (2.938E-06)   & 6.09E-03 \, (4.57E-03)\\
$0.15$   & $64^3$  & 1.207E-15 \, (1.294E-08)   & 7.13E-02 \, (5.83E-02)\\
$0.15$   & $128^3$ & 1.319E-15 \, (1.317E-08)   & 6.79E-01 \, (5.49E-01)\\\hline

$0.10$   & $16^3$  & 3.802E-01 \, (3.802E-01)   & 8.25E-04 \, (6.86E-04)\\
$0.10$   & $32^3$  & 1.129E-03 \, (1.129E-03)   & 5.47E-03 \, (4.54E-03)\\
$0.10$   & $64^3$  & 2.624E-09 \, (2.904E-08)   & 7.07E-02 \, (5.83E-02)\\
$0.10$   & $128^3$ & 1.682E-15 \, (2.943E-08)   & 7.12E-01 \, (5.60E-01)\\
\bottomrule
\end{tabular}
\end{table}

\begin{figure}
\center
\includegraphics[width=0.75\textwidth]{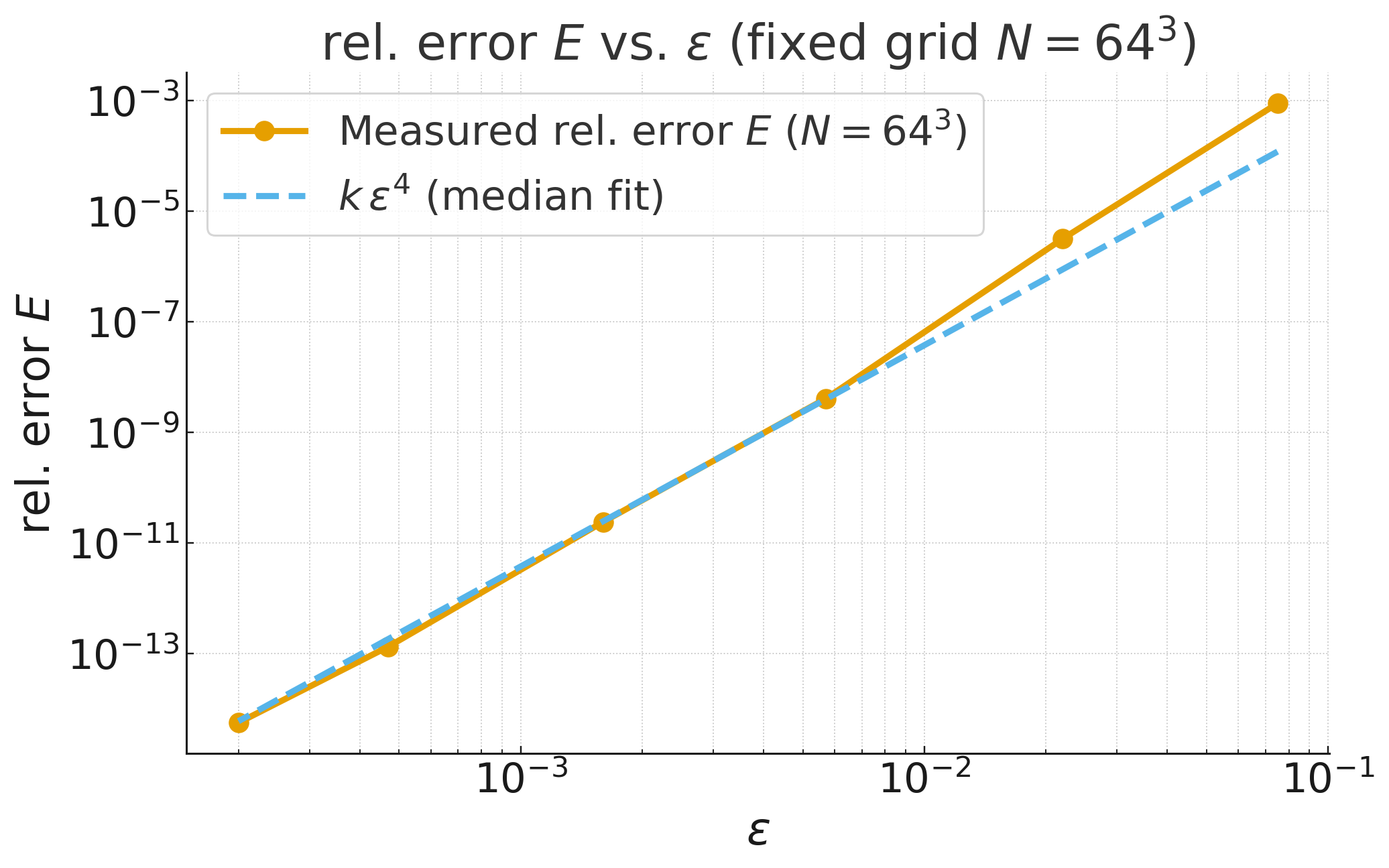}
\caption{\label{fig:eps_convergence}For fixed grid $N=64^3$, relative error $E$ vs. cutoff $\varepsilon$, exhibits a fourth-order decay; a scaled $\varepsilon^4$ reference is overlaid.}
\end{figure}

\change{Theorem~\ref{thm:u-bound} shows that our methods exhibits an error of $\mathcal{O}(\varepsilon^4)$, which eliminates the use of a near-field correction if $\varepsilon$ is chosen small enough. This fourth-order error asymptotics of our proposed method can be seen in the results of Tab.~\ref{tab:vareps}, where we vary $\varepsilon$ for the Gaussian source with $\sigma=0.20$. Fig.~\ref{fig:eps_convergence} shows relative error $E$ versus $\varepsilon$ for a fixed grid $N=64^3$; the measurements follow a reference $\propto \varepsilon^4$ line.}

\begin{table}[h!]
\caption{\change{Errors $E$ for the Gaussian density \eqref{eqn:gaussian} in $[-2,2]^3$ with $\sigma=0.20$ and varying near-field cutoff $\varepsilon$. The last column shows the value of $\varepsilon^4$.}} 
\label{tab:vareps}
\centering
\setlength{\tabcolsep}{16pt}
\begin{tabular}{c c c c}
\toprule
$\varepsilon$ & $N$     & $E$ & $\varepsilon^4$ \\ \hline
7.5E-02   & $16^3$  & 1.461E-03     & 3.164E-05 \\      
7.5E-02   & $32^3$  & 7.919E-04     & 3.164E-05 \\
7.5E-02   & $64^3$  & 8.843E-04     & 3.164E-05 \\
7.5E-02   & $128^3$ & 9.245E-04     & 3.164E-05 \\\hline

2.2E-02   & $16^3$  & 1.658E-03     & 2.342E-07 \\      
2.2E-02   & $32^3$  & 2.848E-06     & 2.342E-07 \\
2.2E-02   & $64^3$  & 3.150E-06     & 2.342E-07 \\
2.2E-02   & $128^3$ & 3.172E-06     & 2.342E-07 \\\hline

5.7E-03   & $16^3$  & 1.659E-03     & 1.055E-09\\      
5.7E-03   & $32^3$  & 4.274E-09     & 1.055E-09 \\
5.7E-03   & $64^3$  & 4.081E-09     & 1.055E-09 \\
5.7E-03   & $128^3$ & 4.109E-09     & 1.055E-09 \\\hline

1.6E-03   & $16^3$  & 1.659E-03     & 6.553E-12\\      
1.6E-03   & $32^3$  & 4.155E-09     & 6.553E-12 \\
1.6E-03   & $64^3$  & 2.400E-11     & 6.553E-12 \\
1.6E-03   & $128^3$ & 2.417E-11     & 6.553E-12 \\\hline

4.7E-04   & $16^3$  & 1.659E-03     & 4.879E-14\\      
4.7E-04   & $32^3$  & 4.154E-09     & 4.879E-14 \\
4.7E-04   & $64^3$  & 1.337E-13     & 4.879E-14 \\
4.7E-04   & $128^3$ & 1.347E-13     & 4.879E-14 \\\hline

2.0E-04   & $16^3$  & 1.659E-03     & 1.600E-15\\      
2.0E-04   & $32^3$  & 4.154E-09     & 1.600E-15 \\
2.0E-04   & $64^3$  & 5.644E-15     & 1.600E-15 \\
2.0E-04   & $128^3$ & 9.296E-15     & 1.600E-15 \\\hline
\bottomrule
\end{tabular}
\end{table}

Next, we test our method on a rectangular domain. Tab.~\ref{tab:gaussian_rec} shows results for the computational domain $\boldsymbol{B}=[-3,2] \times [-2,3.5] \times [-1,5]$ and $\sigma=0.2$. 

\begin{table}[h!]
\caption{Errors and timings (s) for Gaussian density \eqref{eqn:gaussian} in $[-3,2] \times [-2,3.5] \times [-1,5]$ and $\sigma=0.2$. \change{$\varepsilon=10^{-4}$.}} 
\label{tab:gaussian_rec}
\centering
\setlength{\tabcolsep}{16pt}
\begin{tabular}{c c c c}
\toprule
$\sigma$ & $N$     & $E$           & $t$ \\ \hline
$0.20$   & $16^3$  & 4.417E-02     & 1.40E-03 \\      
$0.20$   & $32^3$  & 1.857E-05     & 5.02E-03 \\
$0.20$   & $64^3$  & 5.251E-14     & 4.82E-02 \\
$0.20$   & $128^3$ & 4.441E-15     & 5.36E-01 \\
\bottomrule
\end{tabular}
\end{table}

The following validates the method for the two superposed Gaussian densities
\begin{align}\label{eqn:gaussians}
 \rho(\vec x) = \frac{1}{2}\Big( \frac{1}{(2\pi)^{3/2}\sigma_1^3 } e^{-|\vec x-(\vec c+\vec d)|^2/(2\sigma_1^2)} + \frac{1}{(2\pi)^{3/2}\sigma_2^3 } e^{-|\vec x-(\vec c-\vec d)|^2/(2\sigma_2^2)} \Big), 
\end{align}
where $\vec c,\vec d\in \mathbb{R}^3$ are the center of the computational box and a shift, respectively. The exact solution is 
\begin{align}
 u^\ast(\vec x) =  \frac{1}{2}\Big(\frac{1}{4 \pi |\vec x-(\vec c+\vec d)|} \textrm{Erf}\Big(\frac{|\vec x-(\vec c+\vec d)|}{\sqrt{2}\sigma_1}\Big) + \frac{1}{4 \pi |\vec x-(\vec c-\vec d)|} \textrm{Erf}\Big(\frac{|\vec x-(\vec c-\vec d)|}{\sqrt{2}\sigma_2}\Big)\Big). 
\end{align}
The related results are given in Tab.~\ref{tab:gaussians}.

\begin{table}[h!]
\caption{Errors and timings (s) for Gaussian density \eqref{eqn:gaussians} in $[-2,2]^3$ with $\vec d = (0.1,-0.05,0.05)^T$ and $\sigma_1 = 0.2$ and $\sigma_2 = 0.1$. \change{$\varepsilon=10^{-4}$.}} 
\label{tab:gaussians}
\centering
\setlength{\tabcolsep}{16pt}
\begin{tabular}{c c c c}
\toprule
$(\sigma_1,\sigma_2)$ & $N$     & $E$           & $t$ \\ \hline
$(0.20,0.10)$   & $16^3$  & 5.663E-02     & 1.49E-03 \\      
$(0.20,0.10)$   & $32^3$  & 1.533E-03     & 4.86E-03 \\
$(0.20,0.10)$   & $64^3$  & 5.920E-09     & 6.64E-02 \\
$(0.20,0.10)$   & $128^3$ & 6.664E-15     & 5.36E-01 \\
\bottomrule
\end{tabular}
\end{table}

\subsection{Bump function}
Next a bump function is considered with density ($d,R>0$)
\begin{equation}\label{eqn:bump}
 \rho_{R,c}(\vec x) = 
 \left\{\begin{array}{ll}
  2d R^2 \,\frac{3R^4-2R^2 |\vec x-\vec c|^2 - |\vec x-\vec c|^4 - d R^2 |\vec x-\vec c|^2 }{(R^2-|\vec x-\vec c|^2 )^{4}}\,  e^{-d R^2\,(R+|\vec x-\vec c|) / (R - |\vec x-\vec c|)}, & |\vec x-\vec c|<R\\
  0, & |\vec x-\vec c|\geq R,
 \end{array}\right.
\end{equation}
where the exact solution is given as
\begin{align}
 u^\ast(\vec x) = 
 \left\{\begin{array}{ll}
         e^{\tfrac{-d}{1-|\vec x-\vec c|^2 /R^2}} & |\vec x-\vec c|<R\\
         0, & |\vec x-\vec c|\geq R.
        \end{array}\right.
\end{align}

\begin{table}[h!]
\caption{Errors and timings (s) for the bump function density in \eqref{eqn:bump} in $[-3,1] \times [-2,3] \times [-2,4]$ with $d = 10$ and $R = 2$. \change{$\varepsilon=10^{-4}$.}} 
\label{tab:bump}
\centering
\setlength{\tabcolsep}{16pt}
\begin{tabular}{c c c c}
\toprule
$(R,d)$ & $N$     & $E$           & $t$ \\ \hline
$(2,10)$   & $16^3$  & 2.070E-03     & 1.55E-03 \\      
$(2,10)$   & $32^3$  & 3.928E-06     & 5.70E-03 \\
$(2,10)$   & $64^3$  & 9.264E-10     & 5.64E-02 \\
$(2,10)$   & $128^3$ & 1.039E-14     & 5.36E-01 \\
\bottomrule
\end{tabular}
\end{table}

Tab.~\ref{tab:bump} shows the results in the rectangular domain $[-3,1] \times [-2,3] \times [-2,4]$ with $d = 10$ and $R = 2$. We observe improvements in error compared to the results in \cite{exl2017gpu}. 

\subsection{Anisotropic densities}
The following are tests on the anisotropic density ($\vec c = (c_x,c_y,c_z)^T,\,\,\sigma_x,\sigma_y,\sigma_z > 0$)
\begin{align}\label{eqn:anis}
  \rho(\vec x) =  -( \frac{4(x-c_x)^2}{\sigma_x^4} + \frac{4(y-c_y)^2}{\sigma_y^4} + \frac{4(z-c_z)^2}{\sigma_z^4} - \frac{2}{\sigma_x^2}- \frac{2}{\sigma_y^2}- \frac{2}{\sigma_z^2} )\,e^{-(x-c_x)^2/\sigma_x^2 -(y-c_y)^2/\sigma_y^2 -(z-c_z)^2/\sigma_z^2}, 
\end{align}
with the prescribed exact solution 
\begin{align}
 u^\ast(\vec x) = e^{-(x-c_x)^2/\sigma_x^2 -(y-c_y)^2/\sigma_y^2 -(z-c_z)^2/\sigma_z^2}.
\end{align}
Tab.~\ref{tab:anis} shows the results for $\boldsymbol{\sigma} = (0.30,0.20,0.28)^T$ on the domain $\boldsymbol{B} = [-2,2]^3$.\\

\begin{table}[h!]
\caption{Errors and timings (s) for the anisotropic density in \eqref{eqn:anis} in $[-2,2]^3$ with $\boldsymbol{\sigma} = (0.30,0.20,0.28)^T$. \change{$\varepsilon=10^{-4}$.}} 
\label{tab:anis}
\centering
\setlength{\tabcolsep}{16pt}
\begin{tabular}{c c c c}
\toprule
$\boldsymbol{\sigma}$ & $N$     & $E$           & $t$ \\ \hline
$(0.30,0.20,0.28)$   & $16^3$  & 4.208E-01     & 1.32E-03 \\      
$(0.30,0.20,0.28)$   & $32^3$  & 1.627E-04     & 5.62E-03 \\
$(0.30,0.20,0.28)$   & $64^3$  & 1.515E-13     & 5.57E-02 \\
$(0.30,0.20,0.28)$   & $128^3$ & 1.137E-14     & 5.33E-01 \\
\bottomrule
\end{tabular}
\end{table}

Anisotropic densities are known to be difficult cases. Therefore, the next example investigates the behavior for flat domains $\boldsymbol{B} = [-2,2] \times [-2L,2L]^2$ for $L = 2,4,8,16,32$ and adjusted 
$\boldsymbol{\sigma} = (0.20,0.20\,L,0.20\,L)^T$. Results are given in Tab.~\ref{tab:anisL}, which show significantly improved errors compared to \cite{exl2017gpu}, where the respective errors were stagnating above the level of $1$E-$10$ already for the simplest case of $L=2$.

\begin{table}[h!]
\caption{Errors and timings (s) for the anisotropic density in \eqref{eqn:anis} in $[-2,2] \times [-2L,2L]^2$, $L = 2,4,8,16,32$. \change{$\varepsilon=10^{-4}$.}} 
\label{tab:anisL}
\centering
\setlength{\tabcolsep}{12pt}
\begin{tabular}{c c c c}
\toprule
$L$ & $N$ & $E$ & $t$ \\ \hline
2  & $16^3$   & 1.177E+00 & 1.54E-03 \\
2  & $32^3$   & 1.936E-04 & 5.12E-03 \\
2  & $64^3$   & 1.647E-13 & 6.98E-02 \\
2  & $128^3$  & 4.781E-14 & 5.43E-01 \\\hline
4  & $16^3$   & 1.702E+00 & 7.47E-04 \\
4  & $32^3$   & 2.060E-04 & 6.40E-03 \\
4  & $64^3$   & 1.057E-12 & 5.00E-02 \\
4  & $128^3$  & 3.622E-13 & 5.34E-01 \\\hline
8  & $16^3$   & 2.950E+00 & 6.47E-04 \\
8  & $32^3$   & 2.056E-04 & 5.06E-03 \\
8  & $64^3$   & 8.529E-12 & 4.80E-02 \\
8  & $128^3$  & 4.058E-12 & 5.32E-01 \\\hline
16 & $16^3$   & 5.575E+00 & 6.46E-04 \\
16 & $32^3$   & 1.980E-04 & 5.59E-03 \\
16 & $64^3$   & 7.093E-11 & 4.78E-02 \\
16 & $128^3$  & 6.597E-11 & 5.79E-01 \\\hline
32 & $16^3$   & 1.090E+01 & 8.26E-04 \\
32 & $32^3$   & 1.904E-04 & 6.71E-03 \\
32 & $64^3$   & 1.016E-09 & 4.96E-02 \\
32 & $128^3$  & 1.047E-09 & 5.44E-01 \\
\bottomrule
\end{tabular}
\end{table}

\subsection{Oscillating density}
The final test is for the oscillating density ($\sigma,\omega > 0$)
\begin{align}\label{eqn:osci}
  \rho(\vec x) = & \,e^{-|\vec x-\vec c|^2/\sigma^2} \,( 6\omega \sin(\omega |\vec x-\vec c|^2) - 4\omega^2 |\vec x-\vec c|^2 \cos(\omega |\vec x-\vec c|^2))\\
  & \,- (\frac{4|\vec x-\vec c|^2}{\sigma^4} - \frac{6}{\sigma^2})\, e^{-|\vec x-\vec c|^2/\sigma^2}\,\cos(\omega\,|\vec x-\vec c|^2) 
  -  \frac{8\omega |\vec x-\vec c|^2}{\sigma^2}\,e^{-|\vec x-\vec c|^2/\sigma^2}\,\sin(\omega\,|\vec x-\vec c|^2)
  \end{align}
with the prescribed exact solution 
\begin{align}
 u^\ast(\vec x) = e^{-|\vec x-\vec c|^2/\sigma^2}\,\cos(\omega\,|\vec x-\vec c|^2).
\end{align}
Tab.~\ref{tab:osci} shows the results for $\sigma = 0.30$ and $\omega = 20$ on $\boldsymbol{B} = [-2,2]^3$, indicating spectral convergence.

\begin{table}[h!]
\caption{Errors and timings (s) for the oscillating density in \eqref{eqn:osci} in $[-2,2]^3$ with $\sigma = 0.30$ and $\omega = 20$. \change{$\varepsilon=10^{-4}$.}} 
\label{tab:osci}
\centering
\setlength{\tabcolsep}{16pt}
\begin{tabular}{c c c c}
\toprule
$(\sigma,\omega)$ & $N$     & $E$           & $t$ \\ \hline
$(0.30,20)$   & $16^3$  & 6.179E+00     & 1.36E-03 \\      
$(0.30,20)$   & $32^3$  & 7.921E-03     & 6.87E-03 \\
$(0.30,20)$   & $64^3$  & 3.631E-08     & 5.34E-02 \\
$(0.30,20)$   & $128^3$ & 7.727E-14     & 5.37E-01 \\
\bottomrule
\end{tabular}
\end{table}

\newpage
\section{Conclusion}
We have presented a fast and accurate method for solving the unbounded Poisson equation in three dimensions using a spectral convolution approach based on a super-potential formulation. By expressing the solution as the Laplacian of a convolution with the biharmonic Green's function, the method avoids the singularity of the original Green’s function and achieves high-order accuracy without the need for near-field correction terms. The convolution is regularized via a separable Gaussian-sum approximation, enabling efficient precomputation and the use of simple zero-padded FFT with quasi-linear overall complexity. Unlike earlier Gaussian-sum methods, which require a Taylor-based correction to mitigate cutoff errors, our approach achieves machine-precision accuracy solely from the convolution and spectral differentiation.

Numerical experiments demonstrate that the proposed method significantly improves upon the original GS approach in both error and runtime performance, particularly in anisotropic regimes. The super-potential formulation leads to fast convergence and reduced computational overhead, making the method especially attractive for repeated evaluations as commonly encountered in physical simulations. These advantages establish the method as a robust and efficient tool for high-fidelity solutions of the free-space Poisson problem on uniform grids.

\section*{Acknowledgements}
\noindent Financial support by the Austrian Science Fund (FWF) via project ”Data-driven Reduced Order Approaches for Micromagnetism (Data-ROAM)” (Grant-DOI: 10.55776/PAT7615923) is gratefully acknowledged. The authors acknowledge the University of Vienna research platform MMM Mathematics - Magnetism - Materials. The computations were partly achieved by using the Vienna Scientific Cluster (VSC) via the funded projects No. 71140 and 71952.
This research was funded in whole or in part by the Austrian Science Fund (FWF) [10.55776/PAT7615923]. For the purpose of Open Access, the author has applied a CC BY public copyright license to any Author Accepted Manuscript (AAM) version arising from this submission. 


\end{document}